\documentclass[a4paper]{article}

\usepackage{icrc2013}
\usepackage[english]{babel}

\title{4$\,$m Davies-Cotton telescope for the Cherenkov Telescope
  Array}

\shorttitle{$4\,$m telescope for the CTA}

\authors{R.~Moderski$^{4}$, J.A.~Aguilar$^{6}$, A.~Barnacka$^{4}$,
  A.~Basili$^{6,9}$, V.~Boccone$^{6}$, L.~Bogacz$^{3}$,
  F.~Cadoux$^{6}$, A.~Christov$^{6}$, M.~Della~Volpe$^{6}$,
  M.~Dyrda$^{2}$, A.~Frankowski$^{4}$, M.~Grudzi{\'n}ska$^{10}$,
  M.~Janiak$^4$, M.~Karczewski$^{5}$, J.~Kasperek$^{1}$,
  W.~Kocha{\'n}ski$^{2}$, P.~Korohoda$^{1}$, J.~Kozio{\l}$^{3}$,
  P.~Lubi{\'n}ski$^{4}$, J.~Ludwin$^{2}$, E.~Lyard$^{8}$,
  A.~Marsza{\l}ek$^{3}$, J.~Micha{\l}owski$^{2}$, T.~Montaruli$^{6}$,
  J.~Nicolau-Kukli{\'n}ski$^{5}$, J.~Niemiec$^{2}$, M.~Ostrowski$^{3}$,
  {\L}.~P{\l}atos$^{5}$, P.J.~Rajda$^{1}$, M.~Rameez$^{6}$,
  W.~Romaszkan$^{1}$, M.~Rupi{\'n}ski$^{1}$, K.~Seweryn$^{5}$,
  M.~Stodulska$^{3}$, M.~Stodulski$^{2}$, R.~Walter$^{8}$,
  K.~Winiarski$^{1}$, {\L}.~Wi{\'s}niewski$^{5}$,
  A.~Zagda{\'n}ski$^{3}$, K.~Zi{\c e}tara$^{3}$,
  P. Zi{\'o}{\l}kowski$^{2}$, P.~{\.Z}ychowski$^{2}$ for the the CTA
  Consortium }

\afiliations{
$^1$ Faculty of Computer Science, Electronics, and Telecommunications, AGH University of Science and Technology, Al.~Mickiewicza 30, 30-059 Cracow, Poland \\
$^2$ Institute of Nuclear Physics, Polish Academy of Sciences, ul.~Radzikowskiego 152, 31-342 Cracow, Poland\\
$^3$ Jagiellonian University, ul.~Orla 171, 30-244 Cracow, Poland \\
$^{4}$ Nicolaus Copernicus Astronomical Center, Polish Academy of Sciences, ul.~Bartycka 18, 00-716 Warsaw, Poland\\
$^{5}$ Space Research Centre, Polish Academy of Sciences, ul.~Bartycka 18A, 00-716 Warsaw, Poland\\
$^{6}$ D\'epartement de physique nucl\'eaire et corpusculaire, Universit\'e de Gen\`eve, CH-1211, Switzerland\\
$^{7}$ ETH Zurich, Inst.\ for Particle Physics, Schafmattstr. 20, CH-8093 Zurich, Switzerland\\
$^{8}$ ISDC Data Centre for Astrophysics, Observatory of Geneva, Universit\'e de Gen\`eve, Chemin d’Ecogia 19, CH-1290 Versoix, Switzerland \\
$^{9}$ Physik Institut, Universit\"at Z\"urich, Winterthurerstr.~190, CH-8057, Switzerland \\
$^{10}$ Astronomical Observatory, Warsaw University, Al.~Ujazdowskie 4, 00-478 Warsaw, Poland
}

\email{moderski@camk.edu.pl}

\abstract{The Cherenkov Telescope Array (CTA) is the next generation
  very high energy gamma-ray observatory.  It will consist of three
  classes of telescopes, of large, medium and small sizes.  The small
  telescopes, of $4\,$m diameter, will be dedicated to the
  observations of the highest energy gamma-rays, above several TeV.
  We present the technical characteristics of a single mirror, $4\,$m
  diameter, Davies-Cotton telescope for the CTA and the performance of
  the sub-array consisting of the telescopes of this type.  The telescope
  will be equipped with a fully digital camera based on custom made,
  hexagonal Geiger-mode avalanche photodiodes.  The development of
  cameras based on such devices is an RnD since traditionally
  photomultipliers are used.  The photodiodes are now being
  characterized at various institutions of the CTA Consortium.  Glass
  mirrors will be used, although an alternative is being considered:
  composite mirrors that could be adopted if they meet the project
  requirements.  We present a design of the telescope structure, its
  components and results of the numerical simulations of the telescope
  performance.}


\begin{document}
\maketitle

\section{Introduction}
The Cherenkov Telescope Array (CTA) will be the next generation
observatory of the very high energy (VHE; $>10\,$GeV) gamma rays
\cite{bib:designcta}.  It will provide unprecedented sensitivity in
the energy range $10\,$GeV-$300\,$TeV.  To achieve its goals the array
will consists of at least tree types of telescopes.  The sensitivity
in the highest energy range, above several TeV, will be provided by
the sub-array of the, so called, ``small size telescopes'' (SSTs).
The SST sub-array is expected to include around $70$ telescopes.

\begin{figure}[h]
  \centering
  \includegraphics[width=0.4\textwidth]{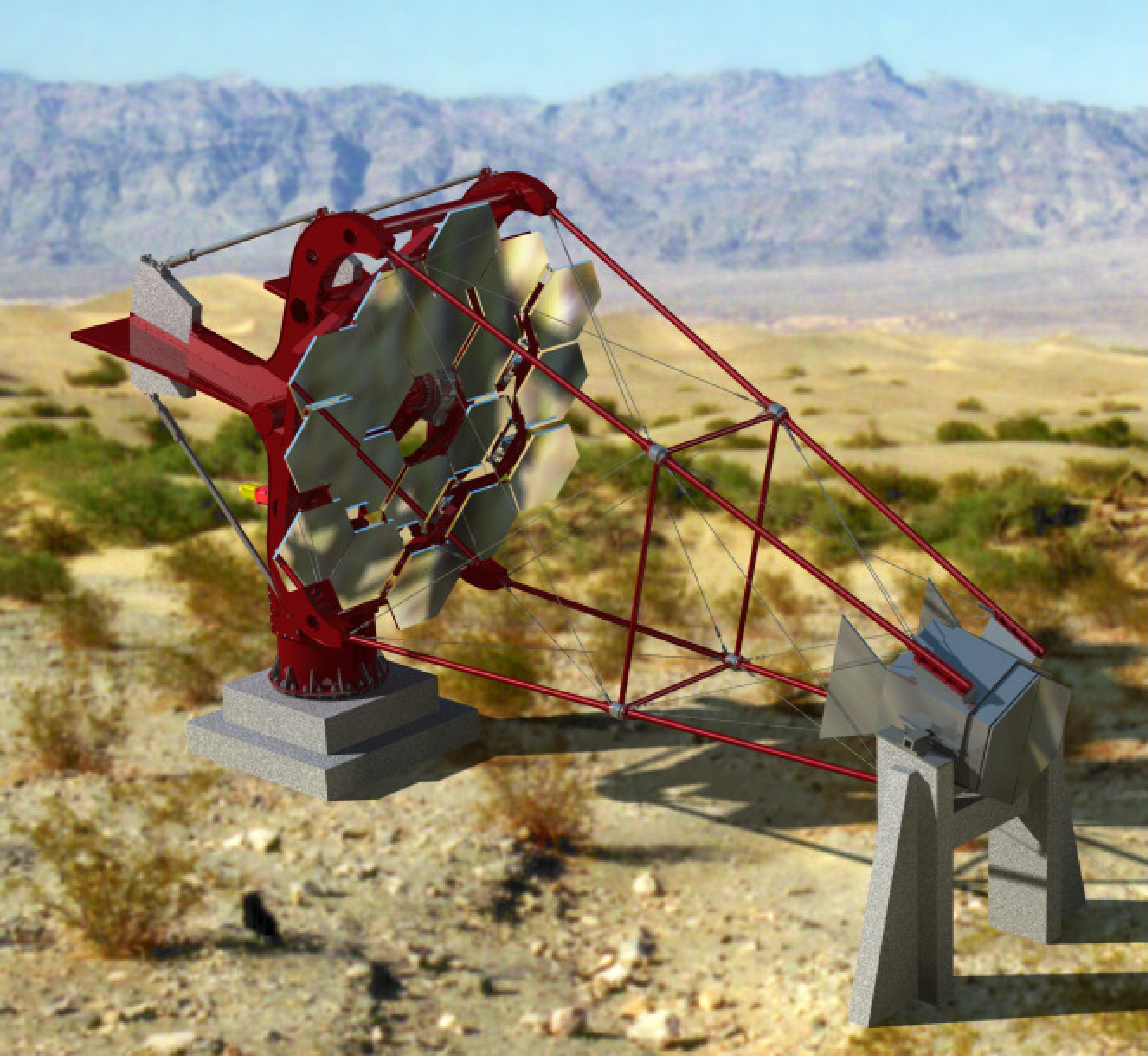}
  \caption{The general view of the $4\,$m Davies-Cotton small size
    telescope based on the technical documentation.}
  \label{fig:telescope}
\end{figure}

A prototype of an SST is currently being developed by a consortium of
Polish and Swiss institutions.  The prototype will be based on the
proven Davies-Cotton (DC) design, used in the currently operated VHE
gamma-ray observatories like H.E.S.S. or VERITAS.  A new idea is to
equip the telescope with a fully digital camera based on silicon
photodetectors.  The overall view of the telescope is presented in
Fig.~\ref{fig:telescope}.

\section{$\mathbf{4}\,$m Davies-Cotton Small Size Telescope}
\subsection{Structure}
The telescope frame and the drive system has been designed at the
Institute of Nuclear Physics, Polish Academy of Sciences (INP) in
Krak\'ow, Poland.  The frame is made of steel.  A camera is placed in
front of the mirror dish on a quadrupod.  The dish is fixed to the
dish support structure, that contains the counterweights, and is
mounted on the telescope support.  The quadrupod is connected to the
dish support in order not to deliver any direct stress on the mirror.

The telescope support consists of a tower fitted with an azimuth drive
system, and a special head with an elevation drive.  Both the azimuth
and the elevation drive systems are based on a set of a roller-bearing
and an IMO slew transmission equipped with two servo-motors.  When in
horizontal position the overall dimensions of the telescope are
roughly $5\,$m height $\times$ $9\,$m long $\times$ $3.5\,$m wide. The
weight of the telescope is around $9\,$t.  For a transportation from a
production site to the assembly point three telescope structures can
be packed in a standard, open-top $12\,$m container.

The telescope structure design has been optimized and checked for
conformance with the CTA specifications through the Finite Element
Method (FEM) analysis.  Such analysis delivers information on the
telescope deformation and mechanical stresses, camera displacement,
and structure eigenfrequencies for various loads, including the
gravity, earthquake, snow and ice, and the wind conditions expected at
a future CTA site.  For the regular observing conditions the maximum
camera displacement with respect to the mirror dish is about $8\,$mm,
which is $1/3$ of the physical pixel size.  The structure is also
strong enough to sustain all extreme load cases -- the mechanical
stresses in the structure are well below the plasticity of the
materials used.  Finally, the lowest eigenfrequencies of the structure
are $3.8\,$Hz, $4.5\,$Hz, and $11.4\,$Hz.  For more details on the
telescope structure see~\cite{bib:niemiec}.

\subsection{Mirror}
The main telescope mirror has a spherical shape and the focal length
of $f = 5.6\,$m.  It consists of $18$ hexagonal facets of $78\,$cm
dimension (flat-to-flat).  The facets are also spherical with a radius
of curvature $R = 2f = 11.2\,$m.  Such a number of facets and the
facet size has been chosen to maximize the mirror area, while keeping
the point spread function (PSF) of the mirror within the required
$0.25^\circ$.  The total mirror area corrected for facets inclination
is $9.42\,\mathrm{m}^2$, while the PSF, determined through
ray-tracing, is $0.21^\circ$ for the rays at an off-axis angle of
$4^\circ$.  A camera housing together with a quadrupod causes a
shadowing of $20\%$ of the light, thus the final collecting area of
the mirror is $7.6\,\mathrm{m}^2$.  The mirror is not isochronous, but
the optical time spread is less than $0.84\,$ns (rms).

Glass mirror facets are foreseen for the telescope.  Glass will be
coated with Al+SiO$_2$+HfO$_2$ coating to maximize the reflectance and
provide weatherability.  The expected average reflectance is of the
order of $94\%$ in the wavelength range $300-550\,$nm.  Alternative
coatings may be considered for the mirror, which include simple
Al+SiO$_2$, or multilayer dielectric coating.

\subsubsection{Alternatives to glass mirrors}
\begin{figure}[h]
  \centering
  \includegraphics[width=0.4\textwidth]{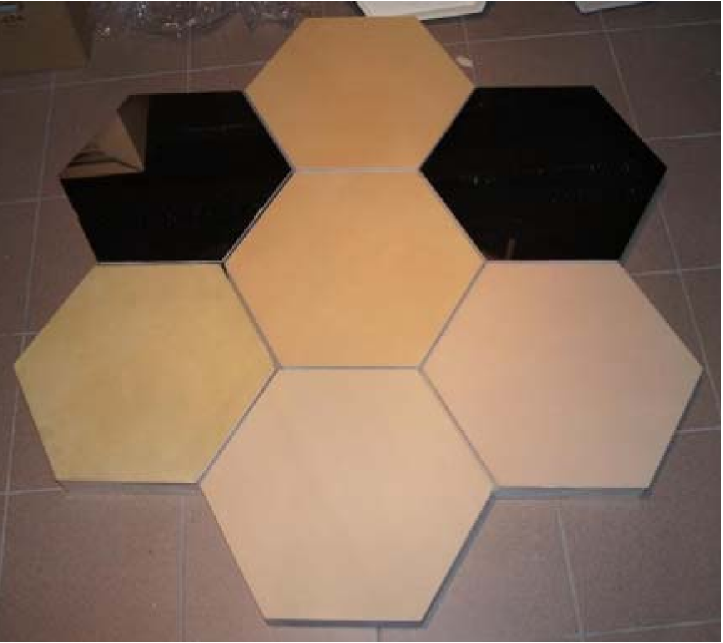}
  \caption{7 composite mirror prototypes (2 coated) developed at the
    Space Research Centre.}
  \label{fig:srcmirror}
\end{figure}

Two alternatives for the glass mirrors are currently being developed
at the Space Research Centre of the Polish Academy of Sciences, Warsaw
(SRC) and INP.  Both technologies use composite materials instead of
glass to speed up the production process and reduce the mirror
mass. SRC technology is based on sheet moulding compound (SMC) -- a
fiber reinforced thermoset material, formed in a high temperature,
high pressure steel mold.  The coating is applied directly on the
composite surface (see Fig.~\ref{fig:srcmirror}).  In the technology
developed by INP an aluminum V-shaped honeycomb structure is used to
support two glass panels and additional $1\,$mm thick glass sheet is
cold slumped on the spherical layer of epoxy resin.  Deposition of
resin layer and cold slumping is performed on the mould.  Both
technologies are now intensively tested to prove the fulfillment of
the requirements.

\subsubsection{Mirror adjustment system}
\begin{figure}[h]
  \centering
  \includegraphics[width=0.4\textwidth]{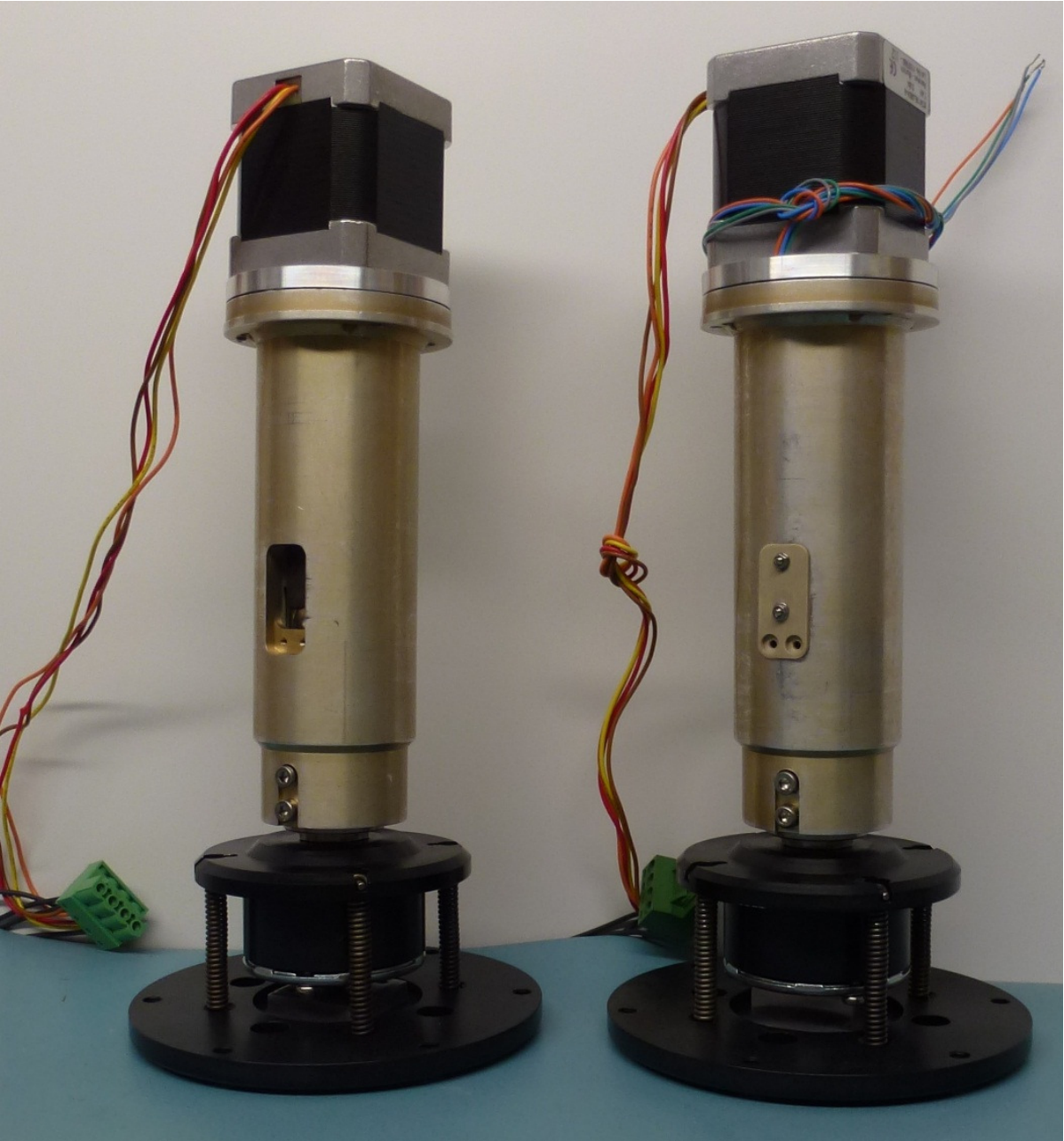}
  \caption{Two prototype actuators build at the Space Research
    Centre.}
  \label{fig:actuatorsrc}
\end{figure}

Each mirror facet is going to be equipped with the justification
system to allow focusing of the whole mirror.  For each facet the
system consists of tree actuators: one fixed, one movable with one
degree of freedom, and one movable with two degrees of freedom.
Control electronics completes the system. To reduce the number of
cables the movable actuators are going to be controlled wirelessly.
The focusing is performed by observing the image of stars on the lid
of the main telescope camera with a dedicated CCD camera.  Three
versions of the system are currently being developed by
Universit{\"a}t Z{\"u}rich, SRC (see Fig.~\ref{fig:actuatorsrc}), and
Universit{\"a}t T{\"u}bingen.  The system will be able to position the
mirror facets with $2\,\mu$m accuracy.

\subsection{Camera}
The main telescope camera is based on the FlashCam concept
\cite{bib:flashcam} and bears the working name DigiCam. The design
separates the photon detection plane (PDP) from the camera electronics
thus allowing these two parts to be physically placed in different
locations, e.g.\ PDP in the focal plane of the telescope and camera
electronics in a box outside of the telescope structure.  In the
camera the signal coming from the photodetectors, after amplification
and possibly shaping, is digitized and both trigger decisions and
readout is done on digital signal.  Such a scheme allows for a great
flexibility in trigger algorithms and readout organization.

Contrary to the original FlashCam design DigiCam will use the Geiger
avalanche photodiodes (G-APD) instead of the vacuum tube
photomultipliers (PMT) \cite{bib:gapd}.  These are new semiconductor
photodetectors used with a great success in the FACT camera
\cite{bib:fact}.  They allow for a smaller size of the PDP, do not
require a high voltage, and offer very good photon resolution.
G-APDs, contrary to PMT, do not suffer from aging, thus allowing for
observations being performed during moonlight.

The DigiCam camera will consist of $1296$ pixels organized in $108$
modules of $12$ pixels.  With an angular pixel size of $0.25^\circ$
the whole camera will offer $9^\circ$ field of view.  The total camera
weight is expected to be about $300\,$kg and should fit into a
cylinder of dimensions $1.2\,$m diameter $\times$ $1\,$m length.

The DigiCam focal plane is developed by Universit{\'e} de Geneve and
Universit\"at Z\"urich.  For the readout electronics, the FlashCam
readout system (see~\cite{bib:flashcam}) will be adopted.

\subsubsection{Photon detection plane}
\begin{figure}[h]
  \centering
  \includegraphics[width=0.4\textwidth]{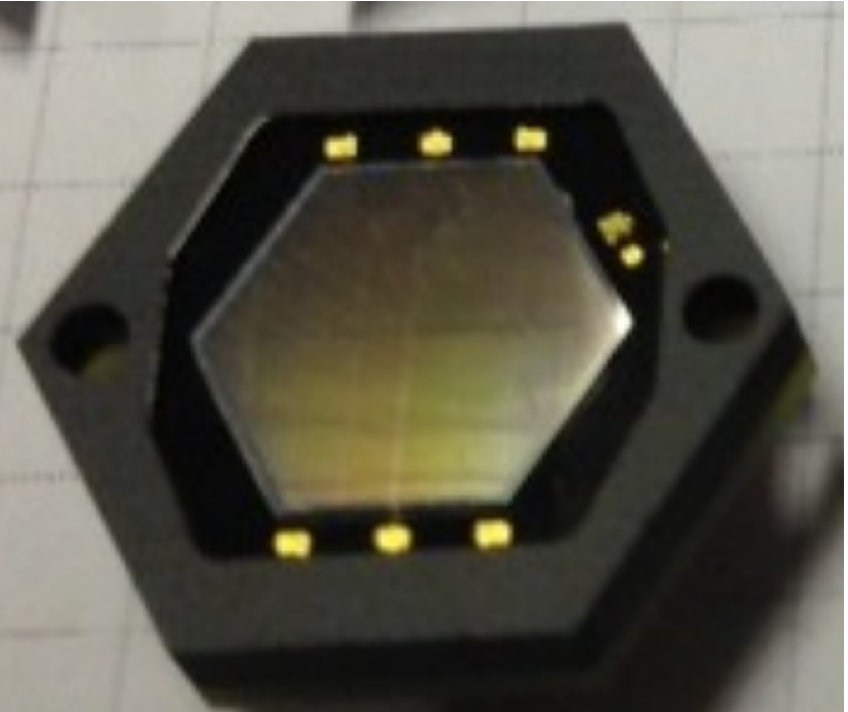}
  \caption{Hamamatsu S12516 Geiger-mode avalanche photodiode.}
  \label{fig:s12516}
\end{figure}

The heart of the PDP is the Hamamtsu S12516 detector (see
Fig.~\ref{fig:s12516}). It is a four-channel, hexagonal shape G-APD
with $50\,\mu$m or $100\,\mu$m cell size, $94\,\mathrm{mm}^2$ total
area offering $35\% - 40\%$ photon detection efficiency at the
supplied voltage of $70\,$V.  Few prototypes of this new G-APD are
under study (see~\cite{bib:boccone}) and the first measurement seems
to confirm the performance to be what was expected.  The devices also
show a reasonable dark current rate and cross-talk despite their large
area.  $12$ detectors will be mounted on single module board, which
will also host the preamplifier, slow control system and power
supply. $108$ of such boards will constitute the whole PDP.

A mosaic of light concentrators will be placed in front of the
photodetectors to reduce the dead space and cut off background light
not coming from the mirror.  After analysis an empty,
hexagonal cone has been chosen with the entrance size of $23.2\,$mm
and compression factor of $6$.  The shape of the cone has been
optimized using Bezier curves and the cut-off angle has been
determined to be $24^\circ$.  Prototype light concentrators have been
fabricated at the company INITIAL and an optimal coating is under
development in collaboration with Thin Film in Z{\"u}rich.

Light concentrators will be covered with an entrance window to provide
protection against water and dust.  The window will be made of
BOROFLOAT\textregistered~33 borosilicate glass from SCHOTT.  The
window can be coated with a dichroic filter to reduce the amount of
nigh sky background light entering the PDP.  The optimal wavelengths
of the filter are under investigation.  It is under consideration also
the application of an anti-reflective coating.

\subsubsection{Camera electronics}
\begin{figure}[h]
  \centering
  \includegraphics[width=0.4\textwidth]{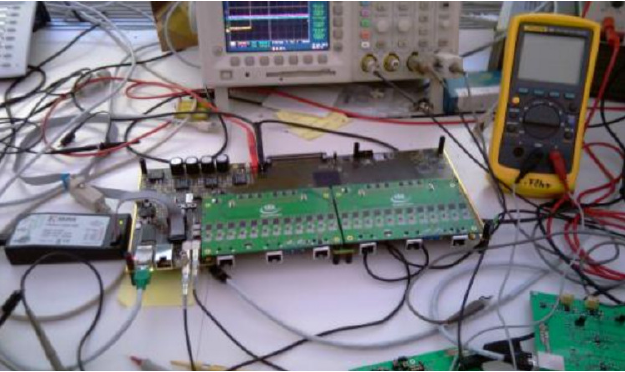}
  \caption{Single motherboard with two FADC cards under tests at the
    Max-Planck Institut f\"ur Kernphysik, Heidelberg.}
  \label{fig:flashmb}
\end{figure}

DigiCam plans to adopt the readout system of the FlashCam.  Analog
signals from the PDP will be transferred via CATx cable to camera
electronic box, where they are digitized using $12\,$ bit, $250\,$MS/s
TI ADS41B29 FADCs.  $12$ of such converters are placed on a single
``FADC board'', and two of such boards can be mounted on a single
motherboard (see Fig.~\ref{fig:flashmb}).  Each motherboard will host
a FPGA chip which will be responsible for signal processing.  $54$
motherboards will be required to handle signals from all $1296$
pixels.  They will be located in $9$ crates and connected through a
special backplanes.  A separate set of ``trigger boards'' will be
necessary to collect trigger data from FADC cards and distribute clock
and trigger signals back to FADC cards. Following the central trigger
signal the data will be sent to the central data acquisition system.

The concept of digitization and data transmission has been extensively
tested with demonstrator setups.  It was shown that $2150\,$MB/s
sustained data transfer is possible without packet loss using raw
Ethernet protocol for up to $84$ simulated FADC boards resulting in
the estimated maximal event rate after central trigger of
$22\,$kHz. The performance of standard CAT5/6 cables to transfer
analog signals has been verified. 

\section{Telescope performance}
The expected telescope performance has been estimated through a number
of numerical simulations of signal processing, of a single telescope
and of an array of the telescopes.  All simulations have been
performed for a Hamamatsu S10985 chip as at the time of the data
generation no measurements for the newest S12516 photodetector were
available.  The present measurements (see~\cite{bib:boccone}),
however, show that the real performance of the device are compatible
with the assumed PDE spectrum used in the simulation.  Light
concentrators transmission curve was calculated using ZEMAX
ray-tracing software (see~\cite{bib:boccone}).  The altitude of
$2000\,$m is assumed for the simulations.

Signal reconstruction procedure has been investigated to verify the
accuracy of the estimation of the intensity of the Cherenkov
light.  The simulation procedure involves generation of signal in
photodetector, change of the signal shape by preamplifiers and
filters, signal digitization, and signal reconstruction using
convolution of smoothed signal with its derivative. It was shown that
for sampling rate of $250\,$MS/s, electronic noise rms corresponding
to $15\%$ of single photoelectron (PE) signal, and thermal noise of
$0.6\,$MHz the systematic error on the absolute intensity of the
Cherenkov light is less than $8.5\%$ for a signal amplitudes above
$20\,$PE. This is less than required value of $10\%$ for SST. For
amplitudes larger than $50\,$PE the error is less than $5\%$.  The
estimated error of the photon arrival time is less than $0.6\,$ns.

For a single telescope one of the important parameter is the level of
the night sky background (NSB).  It is the detection rate of
background photons in a single pixel of the camera.  The NSB level
strongly depends on the optical properties of the mirror and the
quantum efficiency of the photon detector. For our telescope the
nominal (dark sky) NSB level has been estimated to be $85\,$MHz, and
$32\,$MHz for Al+SiO$_2$+HfO$_2$ and dielectric coating, respectively.
The low NSB level for dielectric coating is caused by negligable
reflectance of this coating beyond $550\,$nm, where the NSB intensity
increases significantly.  The same effect can be achieved for
Al+SiO$_2$+HfO$_2$ coating by introduction of a filter in the optical
path of the telescope.  The expected NSB level for partial moonlight
conditions is five times higher than the nominal level.

\begin{figure}[h]
  \centering
  \includegraphics[width=0.4\textwidth]{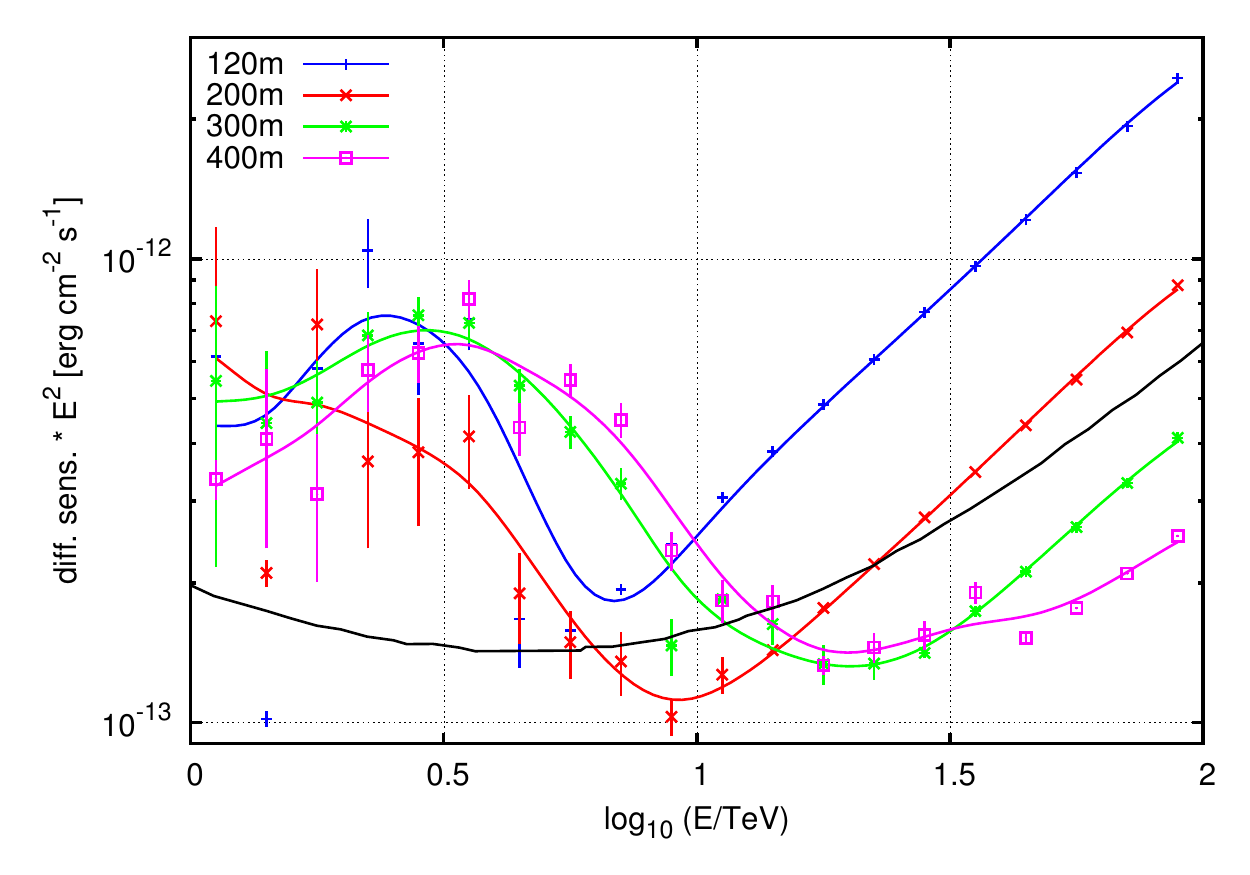}
  \caption{The expected point source differential sensitivity of the
    $64$-telescope array for four different telescope separations:
    $120\,$m, $200\,$m, $300\,$m, and $400\,$m.  $50\,$h observation
    time and $5\,\sigma$ detection significance is assumed.  The
    solid, black line indicates the required sensitivity of the array
    as defined by the CTA project.}
  \label{fig:sens}
\end{figure}

To estimate the performance of the array of the telescopes we use a
concept of ``telescope cell'' \cite{bib:cell}.  In this approach the
cell is a set of four telescopes and only events detected inside the
cell are considered.  The results of the analysis are then
extrapolated to any size of the array.  In such a way lower limits on
the estimated parameters of the array are obtained since events
detected outside of the cell are not taken into account.  The relative
error is smaller if the number of cells becomes larger.  It is worth
noting that to achieve the required $7\,\mathrm{km}^2$ effective area
of the array with less than $70$ telescopes the telescopes need to be
separated by at least $360\,$m.  In our analysis we simulated
different cell sizes, from $120\,$m up to $1000\,$m.  Here we preset
the results for $120\,$m, $200\,$m, $300\,$m, and $400\,$m
separations, since the results for larger distances are significantly
worse.  The obtained differential sensitivity of the $64$-telescope
array is presented in Fig.~\ref{fig:sens}.  The results below
$10\,$TeV are quite uncertain due to low proton statistic.  The
angular resolution of the sub-array is below $0.2^\circ$, but will
improve once the photon arrival time is used in the analysis.  The
details of the analysis are provided in \cite{bib:barnacka}.  It is
clear that such an array is able to fulfill the requirements for a
telescope separation larger than $300\,$m.

The full simulation of the whole $70$-telescope array has been
recently performed with the use of the European grid infrastructure
(see~\cite{bib:prod2}).  The results of these simulations are
currently under investigation.

\section{Conclusion}
The $4\,$m Davies-Cotton telescope equipped with a fully digital
camera meets all requirements of the CTA project.  This telescope has
many advantages: it is lightweight, easy to transport and install,
easy to operate.  Moreover, the estimated cost of $600\,$kEUR of the
prototype indicates that by mass production technique the target cost
of $420\,$kEUR can be reached.

\vspace*{0.5cm} \footnotesize{{\bf Acknowledgment:}{We gratefully
    acknowledge support from the agencies and organizations listed in
    this page: http://www.cta-observatory.org/?q=node/22.  The work
    has been supported by the Swiss National Science Foundation, the
    NCBiR through project ERA-NET-ASPERA/01/10, the NCN through
    project DEC-2011/01/M/ST9/01891, and MNiSzW project
    498/FNiTP/158/2010.  This research was also supported in part by
    PL-Grid infrastructure.}}

\end{document}